\documentclass{PoS}
\usepackage{graphicx,multirow,subfigure}
\usepackage{amssymb}
\usepackage{amsmath}
\newcommand{\tabspace}{0.1} 

\title{Naturalness and Dark Matter Properties of the BLSSM}

\ShortTitle{Naturalness/DM in BLSSM}

\author{Luigi Delle Rose\\
        Particle Physics Department, Rutherford Appleton Laboratory, Chilton, Didcot, Oxon OX11 0QX, United Kingdom\\
        E-mail: \email{L.Delle-Rose@soton.ac.uk}}
    
\author{Shaaban Khalil\\
	Center for Fundamental Physics, Zewail City of Science and Technology, Sheikh Zayed,12588 Giza, Egypt\\
	E-mail: \email{Skhalil@zewailcity.edu.eg}}

\author{\speaker{Simon King}\\
	School of Physics and Astronomy, University of Southampton, Highfield, Southampton SO17 1BJ, United Kingdom\\
	E-mail: \email{sjd.king@soton.ac.uk}}

\author{Carlo Marzo\\
	National Institute of Chemical Physics and Biophysics, R{\"a}vala 10, 10143 Tallinn, Estonia\\
	E-mail: \email{Carlo.Marzo@kbfi.ee}}

\author{Stefano Moretti\\
	School of Physics and Astronomy, University of Southampton, Highfield, Southampton SO17 1BJ, United Kingdom\\
	E-mail: \email{S.Moretti@soton.ac.uk}}

\author{Cem S. Un\\
	Department of Physics, Uluda\~{g} University, TR16059 Bursa, Turkey\\
	E-mail: \email{cemsalihun@uludag.edu.tr}}

\abstract{In this report, we compare the naturalness and Dark Matter (DM) properties of the Minimal Supersymmetric Standard Model (MSSM) and the $B-L$ Supersymmetric Standard Model (BLSSM), with universality in both cases. We do this by adopting standard measures for the quantitative analysis of the Fine-Tuning (FT), at both low (i.e. supersymmetric (SUSY)) and high (i.e. unification) scales. We will see a similar level of FT for both models in these scenarios, with a slightly better FT for the BLSSM at SUSY scales and	MSSM at Grand Unification Theory (GUT) scales. When including DM relic constraints, we drastically confine the MSSM's parameter space, whereas we still find a large parameter space available for the non-minimal scenario.}

\FullConference{XXV International Workshop on Deep-Inelastic Scattering and Related Subjects\\
		3-7 April 2017\\
		University of Birmingham, UK}

\begin{document}

\section{Introduction}
There are two main problems with the Standard Model (SM) that can be addressed with low scale SUSY. Firstly, the hierarchy problem, exemplified by huge loop corrections to the Higgs mass, such that the bare coupling must be fine tuned to $m_H ^2 / \Lambda _{NP} ^2$, where $\Lambda_{NP}$ is the scale of new physics, which would be $\approx$ 1 part in $10^{28}$, if this is taken to be the GUT scale. SUSY partners to the SM particles cancel out these contributions, but since we do not see low scale SUSY partners, but it is broken at some scale, we are left with a FT of the scale $m_H ^2 / \Lambda_{SUSY}^2$. Secondly, there is a huge amount of evidence for DM, which cannot be explained by any particle in the SM. In the MSSM, if one imposes R-parity conservation, the Lightest SUSY Particle (LSP) (the neutralino) cannot decay and so makes a good DM candidate. However, when combining collider searches and the correct relic density requirement for the LSP, the MSSM's parameter space is severely constrained \cite{Abdallah:2015hza}.

In addition to these two problems, we see that there are light, non-vanishing neutrino masses, which are not accounted for in the MSSM. A simple extension would be to add three Right-Handed (RH) neutrinos and a see-saw mechanism, but there is no direct motivation to add exactly three of these SM singlets. In the BLSSM, we see these can be realised as required following the enforcement of anomaly cancellation condition.

This report is organised as follows: in section \ref{sec:BLSSM}, we review the BLSSM and our specific scenario; in section \ref{sec:Collider}, we discuss the various constraints and numerical work we perform; in section \ref{sec:FT}, we discuss the fine-tuning measures we use; in section \ref{sec:Results}, we present the results of our analysis and, in section \ref{sec:Conclusion}, we conclude.

\section{The $B-L$ Supersymmetric Standard Model}
\label{sec:BLSSM}
In the SM, we see that at Lagrangian level there is an exact global $U(1)_B$ and $U(1)_L$ symmetry. At high scale, due to sphaleron processes, only $B-L$ is conserved. If one were to promote this to a gauge symmetry, the SM gauge group would be extended to $G_{BL}=SU(3)_c \times SU(2)_L \times U(1)_Y \times U(1)_{B-L}$, this can motivate adding exactly three SM singlets. One requires, from the anomaly cancellation condition with three $U(1)_{B-L}$ gauge bosons, that three RH particles be introduced to cancel off the contribution from the Left-Handed (LH) neutrinos. We identify these to be the RH neutrinos and, with these, we can explain the lightness of the LH neutrino masses and their large mixing \cite{Khalil:2006yi}, with a type-I see-saw mechanism.
Within the SUSY version of this scenario, the BLSSM, one gains the same benefits as the MSSM, along with this new explanation for neutrino mass. Another interesting feature of the BLSSM is that R-parity is automatically conserved, rather than being imposed by hand to avoid fast proton decay, as in the MSSM. 

We spontaneously break the $U(1)_{B-L}$ in a similar way to the Higgs mechanism. We introduce two complex scalar fields ($\eta _{1,2}$) with lepton number $\pm2$ to break the $B-L$ symmetry which are singlets under the MSSM gauge group, so Electro-Weak Symmetry Breaking (EWSB) is not spoilt. We may now write the superpotential:
	
\begin{eqnarray*}
		W &=&\mu H_{u}H_{d}+Y_{u}^{ij}Q_{i}H_{u}u^{c}_{j}+Y_{d}^{ij}Q_{i}H_{d}d^{c}_{j}+Y_{e}^{ij}L_{i}H_{d}e^{c}_{j} ~~\big{\}}\text{ MSSM} \nonumber\\
		&+&Y_{\nu}^{ij}L_{i}H_{u}N^{c}_{i} + Y^{ij}_{N}N^{c}_{i}N^{c}_{j}\eta_{1}+\mu^{\prime}\eta_{1}\eta_{2} ~~~~~~~~~~~~~~~~~~~~~\big{\}} \text{ BLSSM-specific}
		\label{superpotential}
\end{eqnarray*}
where the top line is the same as in the MSSM and the bottom shows the BLSSM-specific terms. We now summarise the particle content of the BLSSM.
\begin{table}[h]
	\centering
	\resizebox{0.6\columnwidth}{!}{%
		\begin{tabular}{c | c | c | c | c }
			\multicolumn{2}{c|}{Chiral Superfield} & Spin 0 & Spin 1/2  & $G_{B-L} $\\[\tabspace cm] \hline
			&&&\\[-1em]
			\textcolor{black}{RH Sneutrinos / Neutrinos (x3)} &\textcolor{black}{\(\hat{\nu}\)} & \textcolor{black}{$\tilde{\nu}^* _R$} & \textcolor{black}{$\bar{\nu_R}$}  & \textcolor{black}{(\textbf{1}, \textbf{1}, 0, $\frac{1}{2})$} \\
			
			\textcolor{black}{Bileptons/Bileptinos} &\textcolor{black}{\(\hat{\eta}\)} & \textcolor{black}{$\eta$} & \textcolor{black}{$\tilde{\eta}$}  & \textcolor{black}{(\textbf{1}, \textbf{1}, 0, -1)} \\
			
			&\textcolor{black}{\(\hat{\bar{\eta}}\)} & \textcolor{black}{$\bar{\eta}$} & \textcolor{black}{$\tilde{\bar{\eta}}$}  & \textcolor{black}{(\textbf{1}, \textbf{1}, 0, 1)} \\[0.5em] \hline
			
			\multicolumn{2}{c|}{\phantom{\Large{l}} Vector Superfields \phantom{\huge{l}} } & Spin 1/2 & Spin 1 & $G_{B-L} $\\ \hline	
			\multicolumn{2}{c|}{\color{black} \phantom{\huge{l}} BLino / B' boson \phantom{\huge{l}} } & \color{black} $\tilde{B}^{\prime 0}$ & \color{black} $B^{\prime 0}$ & \color{black}(\color{black}\textbf{1} \color{black}\textbf{1}, \color{black}0, \color{black}{0)}		
		\end{tabular}%
	}
\end{table}

In addition to the RH neutrinos, there are their SUSY partners (the sneutrinos), two complex Higgs singlets (the bileptons) and their SUSY partners (the bileptons); a new $B'$ boson, from the breaking of the $U(1)_{B-L}$, and its SUSY partner (the BLino). We will see that the bileptinos, the BLino and the RH sneutrinos are all valid cold DM candidates, which can satisfy the relic density constraints, in addition to the non-observation via Direct Detection (DD). Due to these new particles, the available parameter space, unlike the MSSM, is huge. One may expect this to be at the price of a high FT, if the $Z'$ associated with the $U(1)_{B-L}$ breaking is to be very massive and we have the soft SUSY breaking terms as universal. However, we will see the FT measures in both the MSSM and BLSSM, in a universal scenario, turn out to be very similar in both cases, for both high (GUT) and low (SUSY) scales. In this work, we use a completely universal scenario with $g_1 = g_2 = g_3 = g_{BL}$ with no gauge-kinetic mixing at GUT scale, $\tilde{g}=0$.

\section{Collider and DM bounds}
\label{sec:Collider}
We use the \texttt{SARAH} and \texttt{SPheno} programs \cite{Staub:2013tta,Porod:2003um} and GUT parameters as inputs in producing our spectra. In order for the spectra to be realistically allowed, we place the requirement that the mass of our $Z'$ must satisfy all constraints coming from the LEP2 data, from Electro-Weak Precision Observables (EWPOs) and from Run 2 of the LHC, with $\sqrt{s}=13$ TeV and an integrated luminosity of $\mathcal{L}=13.3$ fb$^{-1}$ as presented in \cite{Accomando:2016sge}. We find that the value of $M_{Z'}=4$ TeV will allow us to safely evade all searches, for all couplings and widths, so we adopt this as the default $Z'$ mass value. We also use the \texttt{HiggsBounds/HiggsSignals} \cite{Bechtle:2008jh,Bechtle:2011sb,Bechtle:2013wla,Bechtle:2015pma,Bechtle:2013xfa} programs, in order to fix that the lightest of our Higgs particles be SM like and that the others must satisfy the non-observations of heavy scalars at the LHC.

In order to find the relic density for each of our spectrum points, we use the \texttt{MicrOMEGAs} program \cite{Belanger:2006is,Belanger:2013oya}. We compare this calculated value to the current measured value of the DM relic density:
\begin{equation} \label{PLANCK}
\Omega h^2 = 0.1187 \pm 0.0017({\rm stat}) \pm 0.0120({\rm syst}) 
\end{equation}
as measured in 2015, by the Planck Collaboration \cite{Ade:2015xua}.

\section{Fine-Tuning Measures}
\label{sec:FT}
In this work, we adopt the most common definition of FT, which is based upon the change in the $Z$-boson mass when altering a fundamental parameter of the theory. Its measure (denoted by $\Delta$) equals the largest of these changes defined as \cite{Ellis:1985yc,Barbieri:1987fn}
\begin{equation}
\Delta={\rm Max} \left| \frac{\partial \ln v^2}{\partial \ln a_i}\right| =  {\rm Max} \left| \frac{a_i}{v^2} \frac{\partial v^2}{\partial a_i } \right|  = {\rm Max} \left| \frac{a_i}{M_Z ^2} \frac{\partial M_Z ^2}{\partial a_i} \right|.
\label{eq:BGFT}
\end{equation}
We will calculate the FT at both low (SUSY) and high (GUT) scale. Calculating the FT in both these regimes will allow us to differentiate if one model is only better fine-tuned at one scale. For the GUT-FT, our high-scale parameters are: the unification masses for scalars ($m_0$)  and gauginos ($m_{1/2}$), the universal trilinear coupling ($A_0$), the $\mu$ parameter and the quadratic soft SUSY term ($B\mu$), for the MSSM. For the BLSSM we have two additional parameters: a $B-L$ version of $\mu$ ($\mu '$) and the corresponding quadratic soft SUSY term, $B \mu '$. So,
\begin{equation}
a_i = \left\lbrace m_0 , ~ m_{1/2},~ A_0,~ \mu ,~ B \mu ,(~\mu' ,~B\mu ') \right\rbrace.
\end{equation}
Recent work \cite{Ross:2017kjc} has shown that loop corrections may affect the absolute scale of FT. In this work, for the GUT scale, we follow the same procedure and eventually find that the total FT will drop by a factor of two for both the MSSM and BLSSM. Now we consider the FT at the SUSY scale, for the MSSM. By minimising the Higgs potential and solving the tadpole equations, one finds the familiar relation for the mass of the $Z$ boson, in terms of SUSY-scale quantities,
\begin{equation}
\frac{1}{2} M_Z^2= \frac{(m_{H_d}^2 + \Sigma_d ) - (m_{H_u}^2 + \Sigma_u) \tan^2 \beta}{\tan^2 \beta -1} -  \mu^2 ,
\label{eq:min_pot_MSSM} 
\end{equation}
where
\begin{equation}
\Sigma_{u,d} = \frac{\partial \Delta V}{\partial v_{u,d} ^2}.
\end{equation}
We consider the loop corrections at the SUSY scale as independent contributions to the $Z$ mass, as in \cite{Baer:2012up}. Substituting this expression into Eq.~(\ref{eq:BGFT}) and using the SUSY-scale parameters $a_i = \lbrace m_{H_d} ^2$, $m_{H_u} ^2$, $\mu ^2$, $\Sigma_u$, $\Sigma _d \rbrace$ for the MSSM, one finds \cite{Baer:2012up}
\begin{equation}
\Delta_{\rm SUSY}\equiv {\rm Max}(C_{i})/(M_{Z}^{2}/2)~,
\label{FT}
\end{equation}
where $C_{i}$ is taken to be each of the terms in eq (\ref{eq:min_pot_MSSM}).
A similar procedure may be carried out for the BLSSM, where the equation to determine $Z$ is modified to
\begin{equation}
\frac{Mz^2}{2}=\frac{1}{X}\left( \frac{ m_{H_d}^2 + \Sigma _{d} }{ \left(\tan ^2(\beta
	)-1\right)}-\frac{ (m_{H_u}^2 + \Sigma _u) \tan ^2(\beta )}{
	\left(\tan ^2(\beta )-1\right)} + \frac{\tilde{g} M_{Z'}^2 Y}{4 g_{BL}
}- \mu ^2 \right), \label{eq:blssm_mz}
\end{equation}
where 
\begin{equation}
X= 1 + \frac{\tilde{g}^{2}}{(g_{1}^{2}+g_{2}^{2})}+\frac{\tilde{g}^{3}Y}{2g_{BL}(g_{1}^{2}+g_{2}^{2})}
\end{equation}
and
\begin{equation}
Y= \frac{\cos(2\beta ')}{\cos (2\beta)} = \frac{\left(\tan^2 {\beta} +1\right) \left(1-\tan^2 {\beta '} \right)}{\left(1-\tan ^2 {\beta } \right) \left(\tan ^2 {\beta '}
	+1\right) }.
\end{equation}
In the limit of no gauge kinetic mixing ($\tilde{g}\rightarrow 0$), this equation reproduces the MSSM minimised potential 
of Eq. (\ref{eq:min_pot_MSSM}).
Our SUSY-FT parameters ($C_i$) for the BLSSM are taken to be each of the terms in eq (\ref{eq:blssm_mz}).
One could expect now that our $M_{Z'}$ is large and so dominates the FT, but the gauge kinetic mixing $\tilde{g}$ will run from 0 at the GUT scale to $\approx 0.1$ at the SUSY scale, so this term will be suppressed. In all four scenarios (SUSY/GUT for MSSM/BLSSM), the dominant parameter to determine the FT is the $\mu$ parameter. The second largest contribution is from $M_{1/2}$, for the GUT scale measures, while in the SUSY case all other parameters contribute negligibly.

\section{Results}
\label{sec:Results}
\begin{figure}[ht!]
	\newcommand{\size}{0.50}
	\subfigure[BLSSM GUT-FT.]{\includegraphics[scale=\size]{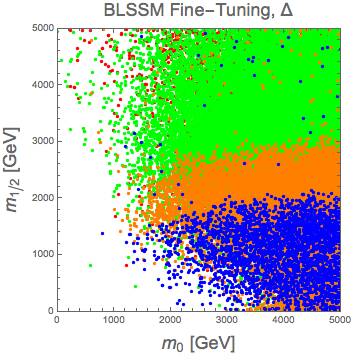} \label{fig:ftbg_blssm_m0_m12}}	
	\subfigure[BLSSM SUSY-FT. ]{\includegraphics[scale=\size]{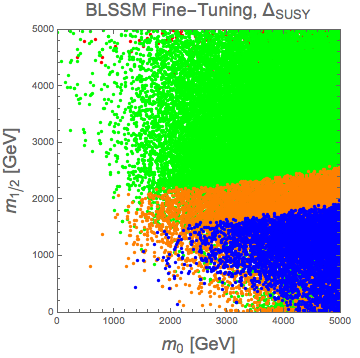}\label{fig:ftew_blssm_m0_m12}}

	\subfigure[MSSM GUT-FT.]{\includegraphics[scale=\size]{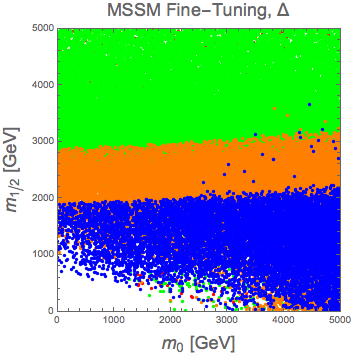}\label{fig:ftbg_mssm_m0_m12}}
	\subfigure[MSSM SUSY-FT. ]{\includegraphics[scale=\size]{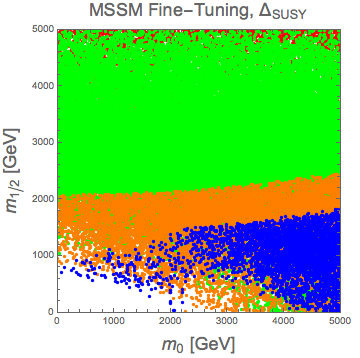}\label{fig:ftew_mssm_m0_m12}}
	\caption{Fine-tuning in the plane of unification of scalar, gaugino masses for BLSSM and MSSM for both GUT-parameters ($\Delta$) and SUSY parameters ($\Delta_{\rm SUSY}$). The FT is indicated by the colour of the dots: blue for FT $<$ 500; Orange for 500 $<$ FT $<$ 1000; Green for 1000 $<$ FT $<$ 5000; and Red for FT $>$ 5000.}
	\label{fig:all_m0_m12}
\end{figure}

We now present the results for the comparison of the FT for SUSY/GUT scales, for the BLSSM/MSSM. In addition we will show how the DM candidates compare for the two models. The scan has been performed over the ranges: [0,5] TeV for $m_0$ and $m_{1/2}$, [0,60] for $\tan \beta$, [-15,15] TeV for $A_0$ and, for the BLSSM alone, [0,2] for $\tan \beta '$ and $[0,1]$ for the Yukawa couplings $Y^{(1,1)}$, $Y^{(2,2)}$, $Y^{(3,3)}$, with $M_{Z'}=4$TeV. For both the MSSM and BLSSM, we required 60,000 data points that satisfied the \texttt{HiggsBounds/HiggsSignals} requirements. Figure \ref{fig:all_m0_m12} shows how the FT compares in the MSSM and BLSSM for low (SUSY) and high (GUT) scale. The spectrum points have been plotted in the plane of the universal scalar and g\textit{}augino masses ($m_0$ and $m_{1/2}$) and coloured according to the values of their FT measure, red for FT $>$ 5000, green for 1000 $<$ FT $<$ 5000, orange for 500 $<$ FT $<$ 1000 and blue (the least finely-tuned points) for FT $<$ 500. The picture in all four cases is very similar. We see that there is a dependence on $m_{1/2}$, which $\mu$ is strongly related to, and not too much dependence on $m_0$. In the BLSSM, we have fewer points at very low $m_0$, as we have a 4 TeV $Z'$, which requires a stronger running of the Renormalisation Group Equations (RGEs). Both GUT and SUSY scale measures of FT are similar, with a slight improvement in the BLSSM, and a worsening in the MSSM.

\begin{figure}[h!]
	\centering
	\includegraphics[scale=0.4]{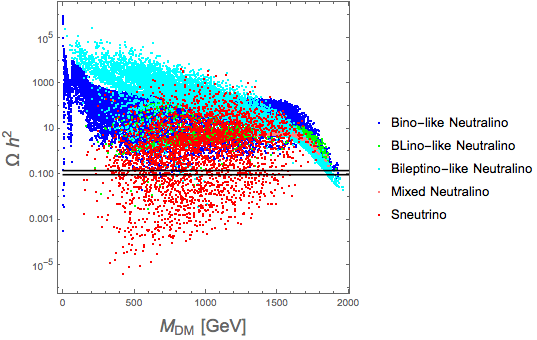}
	\includegraphics[scale=0.4]{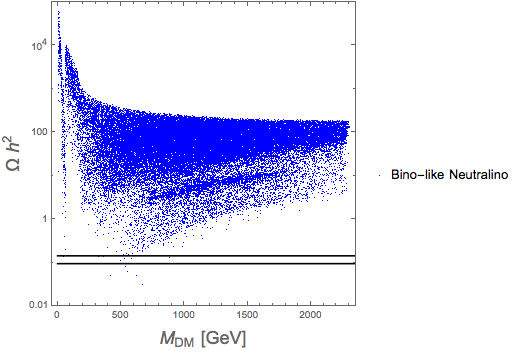}
	\caption{(a) Relic density vs LSP mass for the BLSSM.   
		(b) Relic density vs LSP mass for the MSSM. In both plots the horizontal lines identify the $2\sigma$ region around the current central value of  $\Omega h^2$. }
	\label{fig:BLSSMvsMSSM-DM}	
\end{figure}

We will now turn to the DM sector. For a realistic model of DM, we require that the LSP must comply with cosmological bounds from both DD and relic density. We plot this relic density vs the mass of the DM candidate in figure \ref{fig:BLSSMvsMSSM-DM}. Firstly, in the MSSM there is only one DM candidate, the Bino-like neutralino. However, there are three new candidates for the BLSSM: a BLino and bileptino-like neutralino as well as the lightest RH sneutrino. All of these candidates have points which satisfy the relic density requirement. Secondly, we see that only a tiny fraction of the MSSM Bino candidates can satisfy the relic density requirement (from Planck with a $2 \sigma$ error). However, in the BLSSM we can see that there are many more Bino candidates which satisfy the bounds, in addition to some bileptino and BLino points. The candidate which best satisfies the relic density requirement is the lightest RH sneutrino, since the bounds go directly through the middle of the sneutrino's relic density range.

\section{Conclusions}
\label{sec:Conclusion}
We have compared the FT response and DM sectors of the MSSM and BLSSM, both with universality conditions. We see that the FT is similar in both cases but, once DM constraints are taken into account, the MSSM becomes highly constrained unlike the BLSSM which still has a large available parameter space.

\section*{Acknowledgements}
\noindent
SM is supported in part through the NExT Institute. The work of LDR has been supported by the ``Angelo Della Riccia'' foundation and the STFC/COFUND Rutherford International Fellowship scheme. The work of CM is supported by the ``Angelo Della Riccia'' foundation and by the Centre of Excellence project No TK133 ``Dark Side of the Universe''. The work of SK is partially supported by the STDF project 13858. All authors acknowledge support from the grant H2020-MSCA-RISE-2014 n. 645722 (NonMinimalHiggs).

\end{document}